\documentclass[twocolumn, showpacs,amsmath,amssymb,nofootinbib]{revtex4}

\usepackage{graphicx}
\usepackage{dcolumn}
\usepackage{bm}
\usepackage{epsfig}
\usepackage{color}

\usepackage{color}
\usepackage{hyperref}

\hypersetup{
    colorlinks=true,
    linkcolor=red,
    citecolor=blue,
}

\def\fun#1#2{\lower3.6pt\vbox{\baselineskip0pt\lineskip.9pt
  \ialign{$\mathsurround=0pt#1\hfil##\hfil$\crcr#2\crcr\sim\crcr}}}
\def\simgt{\mathrel{\lower0.6ex\hbox{$\buildrel {\textstyle >}
 \over {\scriptstyle \sim}$}}}
\def\simlt{\mathrel{\lower0.6ex\hbox{$\buildrel {\textstyle <}
 \over {\scriptstyle \sim}$}}}

\def\bea{\begin{eqnarray}}
\def\eea{\end{eqnarray}}
\def\be{\begin{equation}}
\def\ee{\end{equation}}

\input epsf

\def\be{\begin{equation}}
\def\ee{\end{equation}}
\def\ba{\begin{eqnarray}}
\def\ea{\end{eqnarray}}

\newcommand*{\D}{{\rm d}}
\newcommand*{\mpl}{M_{\rm Pl}}

\begin{document}

\preprint{}

\title{Effective theory for the Vainshtein mechanism from the Horndeski action}

\author{Kazuya Koyama$^{(1)}$\email{Kazuya.Koyama@port.ac.uk}, Gustavo Niz$^{(1,2)}$, Gianmassimo Tasinato$^{(1)}$}

\bigskip

\vskip1cm

\affiliation{
\vskip0.3cm
$^{(1)}$Institute of Cosmology \& Gravitation, University of Portsmouth, Dennis Sciama Building, Portsmouth, PO1 3FX, United Kingdom \\
$^{(2)}$ Departamento de F\'{\i}sica, Universidad de Guanajuato,\\
 DCI, Campus Le\' on, C.P. 37150, Le\' on, Guanajuato, M\' exico.\\
}


\begin{abstract}
Starting from the general Horndeski action, we derive the most general effective theory
for scalar perturbations around flat space that allows us to screen fifth forces via the Vainshtein mechanism. The effective theory is described by a generalization of the Galileon Lagrangian, which we use to study the stability of spherically symmetric configurations exhibiting the Vainshtein effect. In particular, we discuss the phenomenological consequences of a scalar-tensor coupling that is absent in the standard Galileon Lagrangian. This coupling controls the superluminality and stability of fluctuations inside the Vainshtein radius in a way that depends on the density profile of a matter source. Particularly we find that the vacuum solution is unstable due to this coupling.
 \end{abstract}



\maketitle
\section{Introduction}
The late time acceleration of the Universe has become one of the most challenging problems in theoretical physics. The cosmological constant is the simplest option but its required value is unacceptably small from the point of view of effective
field theory.  Various models have been proposed as alternatives to a pure cosmological constant. Many of them make use of scalar fields to achieve the accelerated expansion. Scalar degrees of freedom also appear in modified gravity models where the acceleration is attributed to the failure of general relativity (GR) on cosmological scales.
However, in general scalars mediate fifth forces  strongly constrained by precision tests of gravity at Solar System scales: any dark energy and modified gravity model involving  scalar fields should accommodate a mechanism to suppress the scalar interaction on small scales,  where stringent constraints on the deviation from GR are applicable.

There has been interesting progress in developing screening mechanisms to suppress scalar interactions on small scales. One of the oldest ideas is the Vainshtein mechanism \cite{Vainshtein}, originally discovered in the context of massive gravity (see Ref.~\cite{Babichev:2013usa} for a review). Massive gravitons have five polarisations instead of the two of GR: the helicity-0 mode couples to matter and modifies the prediction of GR. In a linear approximation,  this helicity-0 mode does not decouple in the massless limit, leading to the so-called vDVZ discontinuity \cite{vanDam, Zakharov}.
This problem can be solved by the Vainshtein mechanism. If the graviton mass is small,
the derivative self-interactions of the helicity-0 mode become important at much larger distances compared with the Schwarzschild radius of a source, and they suppress the coupling of the helicity-0 mode to matter.
The essential feature of this mechanism, which makes it different from other proposals, is that  derivative self-interactions of the scalars are the main factor responsible for the screening mechanism, without requiring any particular form of scalar potential, and any couplings of the scalar to matter other than gravitational.
In the more direct extension of the Fierz-Pauli massive gravity \cite{Fierz} coupled to a source, the non-linear interactions involve higher derivatives in the equations of motion,  or this leads to the Boulware-Deser ghost \cite{Boulware}. The non-linear interactions that give a scalar equation of motion containing only up to second derivatives have been constructed imposing a Galilean symmetry, that allows only a handful of terms in the scalar lagrangian
\cite{Nicolis:2008in}. The covariant version of this theory has been obtained \cite{Deffayet0, Deffayet1} as the most general scalar-tensor theory whose equation of motion contains up to the second derivative of the fields
\cite{Deffayet2, Kobayashi:2011nu}. The action describing this theory was indeed first obtained by Horndeski in 1974 \cite{Horndeski:1974wa} and thus is now called the Horndeski action.

In this paper, we discuss  the general effective theory that describes the Vainshtein mechanism  in  flat space-time starting from the Horndeski action. Although we derive this effective theory from it, the infrared completion of the theory can be described by a different action. Indeed, the effective theory we analyze  includes  suitable  decoupling limits of other scenarios, such as the Dvali-Gabadadze-Porrati (DGP) braneworld model \cite{Dvali:2000hr} or ghost-free massive gravity \cite{deRham:2010ik}.

\section{The Horndeski action}
Our starting point is the Horndeski action \cite{Horndeski:1974wa, Kobayashi:2011nu, Deffayet2}, given by
\begin{eqnarray}\label{Horndeski}
S=S_{\rm GG}\left[g_{\mu\nu}, \phi\right]
+S_{\rm m}\left[g_{\mu\nu},\psi_m\right].
\end{eqnarray}
The first term controls the dynamics of the scalar field, and contains  the following four pieces:
\begin{eqnarray}
S_{\rm GG}=\int\D^4 x\sqrt{-g}\left({\cal L}_2
+{\cal L}_3+{\cal L}_4+{\cal L}_5\right),
\end{eqnarray}
where
\begin{eqnarray}
{\cal L}_2&=&K(\phi, X),
\end{eqnarray}
\begin{eqnarray}
{\cal L}_3&=&-G_3(\phi, X)\Box\phi,\\
{\cal L}_4&=&G_4(\phi, X)R
+G_{4X}(\phi, X)\left[(\Box\phi)^2-(\nabla_\mu\nabla_\nu\phi)^2\right],
\\
{\cal L}_5&=&
G_5(\phi, X)G^{\mu\nu}\nabla_\mu\nabla_\nu\phi
-\frac{1}{6}G_{5X}(\phi, X)\bigl[(\Box\phi)^3
\nonumber\\&&\quad
-3\Box\phi(\nabla_\mu\nabla_\nu\phi)^2
+2(\nabla_\mu\nabla_\nu\phi)^3\bigr],
\end{eqnarray}
with $X:=-g^{\mu\nu}\partial_\mu\phi\partial_\nu\phi/2$, and $(\nabla_\mu\nabla_\nu\phi)^n$ a short notation for the
trace of the $n$-time contraction of $\nabla_\mu\nabla_\nu\phi$.
The notation $A_{B}:=\partial A/\partial B$ will be used throughout the text.
Notice that, in the second part of the action  (\ref{Horndeski}), we are choosing a frame in which matter does not directly couple to the scalar, but only to gravity.

\section{An effective theory for the Vainshtein mechanism}
For the purpose of studying the Vainshtein mechanism, we consider a Minkowski background with a constant scalar field $\phi=\phi_0$ and study the deviations around it, namely
\begin{equation}
\phi = \phi_0 + \pi, \qquad g_{\mu \nu} = \eta_{\mu \nu} + h_{\mu \nu}\,.\label{pertcon}
\end{equation}
The dynamics of fluctuations around this background controls the behavior of gravitational interactions.
Notice that for the background to be a solution, one needs $K(\phi_0)=K_{\phi}(\phi_0)=0$. We construct an effective theory for the dynamics of fluctuations aimed to describe  the  Vainshtein mechanism. As explained above, the distinctive feature of the Vainshtein mechanism is that derivative self-interactions of the scalar are sufficient to control its behavior around a source, and screen its effect. Therefore, in order to identify these derivative interactions and realize the  Vainshtein mechanism as described above, we demand that the scalar does not directly couple to the source, leaving possible couplings to matter through mixing with gravity. In order to do so, we impose that the scalar action respects the Galilean symmetry  $\pi\to\pi+c+b_\mu x^\mu$, introduced in \cite{Nicolis:2008in}. This symmetry prevents direct couplings between the $\pi$ fluctuation and the energy-momentum tensor. Moreover, the symmetry does not allow a potential for the scalar field $\pi$, or couplings between first derivatives of $\pi$ and $\pi$ itself. As a consequence, it is the symmetry that we need in order to isolate the pieces of the action that depend purely on derivative self-interactions of the scalar sector.

We achieve this effective action by expanding the Horndeski action (\ref{Horndeski}) in terms of the fluctuations (\ref{pertcon}), using the following assumptions: the fields $\pi$ and $h_{\mu \nu}$ are small, hence we neglect higher order interactions containing the tensor perturbations $h_{\mu \nu}$, as well as terms containing
higher order powers of the scalar fluctuation $\pi$ and its first derivatives only. On the other hand, we keep all terms with second order derivatives of $\pi$ that preserve the Galilean symmetry, and will provide the necessary self-interactions to realize the Vainshtein screening mechanism. In other words, this means to Taylor-expand the functions $K$, and $G_i$ to first order in the quantities $X$ and $\phi$, and the curvature tensors $G_{\mu\nu}$ and $R$ to first and second order in $h_{\mu\nu}$, respectively: all the higher derivatives of these functions are required to be
suppressed, in order to obtain an effective action that preserves the Galilean symmetry.
By doing this expansion, we schematically find the following terms
\begin{equation}
(\partial h)^2, \quad (\partial \pi)^2, \quad (\partial \pi)^2 (\partial^2 \pi)^n, \quad h (\partial^2 \pi)^n.
\end{equation}
The kinetic term for metric perturbations is contained in $G_4(\phi, x) R$. Thus we introduce the Planck scale as $G_4= \mpl/2$ and canonically normalise $h_{\mu \nu}$ as $\bar{h}_{\mu\nu} = \mpl h_{\mu\nu}$. We further introduce a new mass dimension $\Lambda$ so that terms involving second order derivatives of $\pi$ have the appropriate dimensions. By explicitly expanding the Horndeski action (\ref{Horndeski}) under these prescriptions, we find
\begin{eqnarray}\label{decaction}
{\cal L}^{\rm eff} &=& - \frac{1}{4} \bar{h}^{\mu \nu} {\cal E}^{\alpha \beta}_{\;\;\;\;\; \mu \nu} \bar{h}_{\alpha \beta} + \frac{\eta}{2} \pi \Box \pi  \nonumber\\
&+& \frac{\mu}{\Lambda^3}{\cal L}^{\rm gal}_3
+ \frac{\nu}{\Lambda^6} {\cal L}^{\rm gal}_4 + \frac{\varpi}{\Lambda^9} {\cal L}^{\rm gal}_5 \nonumber\\
&-&  \xi \bar{h}^{\mu \nu} X^{(1)}_{\mu \nu} - \frac{1}{\Lambda^3} \alpha \bar{h}^{\mu \nu}
X^{(2)}_{\mu \nu} + \frac{1}{2 \Lambda^6} \beta \bar{h}^{\mu \nu} X^{(3)}_{\mu \nu}
\nonumber\\
&+& \frac{1}{2 \mpl} \bar{h}^{\mu \nu} T_{\mu \nu},
\end{eqnarray}
where ${\cal L}^{\rm gal}$ are the so-called Galileon terms \cite{Nicolis:2008in}
\begin{eqnarray}
{\cal L}^{\rm gal}_3 &=& -\frac{1}{2} (\partial \pi)^2 [\Pi], \nonumber\\
{\cal L}^{\rm gal}_4 &=& -\frac{1}{2} (\partial \pi)^2 ([\Pi]^2-[\Pi^2]), \nonumber\\
{\cal L}^{\rm gal}_5 &=& -\frac{1}{4} (\partial \pi)^2 ([\Pi]^3-3 [\Pi][\Pi^2] +2 [\Pi^3]),
\end{eqnarray}
and the tensor-scalar couplings $X_i$ are given by
\begin{eqnarray}
X^{(1)}_{\mu \nu} &=& \eta_{\mu \nu} [\Pi] - \Pi_{\mu \nu}, \nonumber\\
X^{(2)}_{\mu \nu} &=& \Pi^2_{\mu \nu} - [\Pi] \Pi_{\mu \nu} +\frac{1}{2} \eta_{\mu \nu}
([\Pi]^2 - [\Pi^2]), \nonumber\\
X^{(3)}_{\mu \nu} &=& 6 \Pi^3_{\mu \nu} - 6 \Pi^2_{\mu \nu} [\Pi] + 3 \Pi_{\mu \nu}
([\Pi]^2 - [\Pi^2]) \nonumber\\
&&  - \eta_{\mu \nu}([\Pi]^3 - 3 [\Pi][\Pi^2]+2 [\Pi^3]),
\end{eqnarray}
where the shorthand notations $\Pi_{\mu \nu}= \partial_{\mu} \partial_{\nu} \pi$, $\Pi^n_{\mu \nu}= \Pi_{\mu \alpha} \Pi^{\alpha \beta}...\Pi_{\lambda \nu}$ and
$[\Pi^n] = \Pi^{n \; \mu}_{\;\;\;\;\;\;\mu}$ are introduced and ${\cal E}^{\alpha \beta}_{\;\;\;\;\; \mu \nu} \bar{h}_{\alpha \beta}$ is the linearised Einstein tensor. We define seven dimensionless parameters $\xi, \eta, \mu, \nu, \varpi,\alpha,\beta$ as follows \cite{Narikawa:2013pjr}
\begin{eqnarray}
G_{4\phi} &=&\mpl \xi, \quad
K_X-2G_{3\phi} =\eta, \nonumber\\
-G_{3X}+3G_{4\phi X} &=& \frac{\mu}{\Lambda^3},
\quad G_{4X}-G_{5\phi} =\frac{\mpl}{\Lambda^3}\alpha,
\nonumber\\
G_{4XX}-\frac{2}{3}G_{5\phi X}&=&\frac{\nu}{\Lambda^6}, \quad
G_{5X}=-\frac{3\mpl}{\Lambda^6}\beta, \nonumber\\
G_{5XX} &=& - \frac{3 \varpi}{\Lambda^9},
\end{eqnarray}
where all functions are evaluated at the background, $\phi=\phi_0$ and $X=0$. We assume these dimensionless parameters are ${\cal O}(1)$ when they are non-zero.

Notice that, as anticipated, the above action respects the Galilean symmetry (up to total derivative terms). On the other hand,
the coupling between the metric and scalar perturbations $\bar{h}^{\mu \nu} X^{(1)}_{\mu \nu}$ and $\bar{h}^{\mu \nu} X^{(2)}_{\mu \nu}$ can be eliminated by the local field redefinition \cite{deRham:2010tw}
\begin{equation}
\bar{h}_{\mu \nu} = \hat{h}_{\mu \nu} - 2 \xi  \pi \eta_{\mu\nu}
+ \frac{2 \alpha}{\Lambda^3} \partial_{\mu} \pi \partial_{\nu} \pi,
\label{trans}
\end{equation}
so that the action (\ref{decaction}) now becomes
\begin{eqnarray}
{\cal L}^{\rm eff} &=& - \frac{1}{4} \hat{h}^{\mu \nu} {\cal E}^{\alpha \beta}_{\;\;\;\;\; \mu \nu} \hat{h}_{\alpha \beta} +
\frac{\eta + 6 \xi^2}{2} \pi \Box \pi \nonumber\\
&+& \frac{\mu + 6 \alpha \xi}{\Lambda^3}{\cal L}^{\rm gal}_3
+
\frac{\nu+2 \alpha^2+4 \beta \xi}{\Lambda^6} {\cal L}^{\rm gal}_4 \nonumber\\
&+& \frac{\varpi+ 10 \alpha \beta}{\Lambda^9} {\cal L}^{\rm gal}_5
+  \frac{1}{2 \Lambda^6} \beta \hat{h}^{\mu \nu} X^{(3)}_{\mu \nu}
\nonumber\\
&+& \frac{1}{2 \mpl} \hat{h}^{\mu \nu} T_{\mu \nu}
- \frac{2 \xi}{\mpl} \pi T + \frac{2 \alpha}{ \mpl \Lambda^3}
\partial_{\mu} \pi \partial_{\nu} \pi T^{\mu \nu}. \nonumber\\
\label{decoupling}
\end{eqnarray}
However, the coupling $\hat{h}^{\mu \nu} X^{(3)}_{\mu \nu}$ cannot be removed by a local field redefinition \cite{deRham:2010tw}. The transformation (\ref{trans}) introduces a coupling between $\pi$ and $T_{\mu\nu}$ via the de-mixing of
scalar from gravity.

This is the most general form of the effective theory aimed to analyse  the Vainshtein mechanism,  which involves up to the second derivatives of the fields in the equations of motion. We should emphasise that, in our discussion,  the scalar field perturbation $\pi$ and its first derivative $\partial \pi$ are assumed to be small. There are  alternative screening mechanisms using the non-linearity of the field itself or the first derivative (see Ref.~\cite{Jain:2010ka} for a review). In these cases, we should expand the functions $K$ and $G_i$
at all orders, hence
 it is required to know the functional forms of these functions. This information is not needed in our approach, which isolates the derivative self-interactions that are needed
to identify and  analyze the Vainshtein mechanism in its traditional form.

As we mentioned before, although we reached this result from the Horndeski action, the infrared completion of the theory can be different from the Horndeski action. For example, the recently proposed massive gravity theory reduces to (\ref{decoupling}) in the decoupling limit with $\eta=\mu=\nu=\varpi=0$ \cite{deRham:2010tw}. In this case, $\Lambda = m^2 \mpl$, with $m$ being the graviton's mass, while $\alpha$ and $\beta$ are related to the two dimensionless parameter of this massive gravity theory, which are usually referred as $\alpha_3$ and $\alpha_4$. A further example is the decoupling limit of the DGP braneworld model, which is reproduced by $\alpha=\beta=\nu=\varpi=0$ \cite{DGPdecouple}. In this case,  $\Lambda = r_c^{-2} \mpl$ where $r_c$ is the cross-over scale between four-dimensional and five-dimensional gravity. In both models, $\pi$ is not a fundamental scalar field, but it is the helicity-0 component of a graviton.

\section{Spherically symmetric solutions}
From the effective action (\ref{decoupling}), it is easy to find static spherically symmetric solutions. We consider the following configuration
\begin{equation}
\hat{h}_{tt}=-2 \Phi, \quad \hat{h}_{ij} = -2 \Psi \delta_{ij},
\quad \phi = \phi_0+\pi(r).
\end{equation}
The field equations yield
\begin{widetext}
\begin{eqnarray}
P(x, A) & \equiv & \xi A(r)+\left(\frac{\eta}{2}+3\xi^2\right)x
+\left(\mu+6\alpha\xi-3\beta A(r)\right)x^2
+\left(\nu+2\alpha^2+4\beta\xi\right)x^3-3\beta^2x^5=0,\\
y(r) &=& \beta x^3+A(r),
\label{basic-equation}
\end{eqnarray}
\end{widetext}
where we define
\begin{eqnarray}
x(r) &=& \frac{1}{\Lambda^3}\frac{\pi'}{r}, \quad
y(r) = \frac{\mpl}{\Lambda^3}\frac{\Phi'}{r} \label{ydef}
 = \frac{\mpl}{\Lambda^3}\frac{\Psi'}{r}, \\
A(r)&=&\frac{1}{\mpl\Lambda^3}\frac{M(r)}{8\pi r^3}.
\end{eqnarray}
The function $M(r)$ represents the mass of a spherically symmetric, pressureless matter source, up to a radius $r$. Outside the
surface $r_s$ of  the  matter source, $A(r)$ can be written as
\begin{equation}
A(r) = \Big( \frac{r_V}{r} \Big)^3, \quad r_V= \Big( \frac{M}{8 \pi \mpl \Lambda^3} \Big)^{1/3},
\end{equation}
where $r_V$ is the Vainshtein radius depending on the total mass $M$ of the source.
Thus, well inside the Vainshtein radius, $A(r) \gg 1$, and the non-linearity of $x$ becomes large. For the Sun, the Vainshtein radius $r_V$ is $0.1$kpc if $\Lambda$ is associated with the energy scale for the current accelerated expansion of the Universe $\Lambda \sim \mpl H_0^2$. Solutions to these equations were studied in \cite{Koyama:prl,Chkareuli:2011te,Sbisa:2012zk} within massive gravity models, and the analysis was extended to the general case in \cite{Kimura:2011dc,Narikawa:2013pjr}.

Stability of these solutions can be studied by considering small perturbations around the background
$ \pi = \pi_0(r) + \varphi(t,r,\Omega), \Phi=\Phi(r) + \delta \Phi(t,r,\Omega),
\Psi=\Psi(r) + \delta \Psi(t,r,\Omega)$ where $\Omega$ represents angular coordinates. By expanding the effective action up to the second order in these small perturbations we find
\begin{equation}
S_{\varphi}= \frac{1}{2} \int d^4 x
\Big[
K_t(r) (\partial_t \varphi)^2 - K_r(r) (\partial_r \varphi)^2
-K_{\Omega}(r) (\partial_{\Omega} \varphi)^2
\Big],
\end{equation}
where
\begin{eqnarray}\label{Ks}
K_r(r) &=& 2 \partial_x P(x,r) |_{x=x_0},
\end{eqnarray}
\begin{eqnarray}
K_t(r) &=& \frac{1}{3 r^2} \frac{d}{dr}
\Big[r^3 \Big\{
\eta+6 \xi^2 + 6 (\mu + 6 \alpha \xi) x \nonumber\\
&+& 12 \alpha A + 18 (\nu + 2 \alpha^2 + 4 \xi) x^2 \nonumber\\
&+& 12 (10 \alpha \beta + \varpi) x^3 + 36 \beta x y
\Big\}
\Big ],\\
K_{\Omega}(r) &=& \frac{1}{2 r} \frac{d}{dr}
\Big[r^2
\Big\{
\eta+6 \xi^2 + 4 (\mu + 6 \alpha \xi) x \nonumber\\
&+& 6 (\nu + 2 \alpha^2 + 4 \xi) x^2 - 12 \beta x y
\Big\}
\Big].
\end{eqnarray}
In addition to the quadratic action for scalar perturbations, there is also a coupling between metric perturbations and the scalar perturbations. The stability of fluctuations depends strongly on the presence of two particular couplings in the Lagrangian (\ref{decoupling}), one given by $\partial_{\mu} \pi \partial_{\nu} \pi T^{\mu \nu}$ and the other by $\hat{h}^{\mu \nu} X^{(3)}_{\mu \nu}$ , which have $\alpha$ and $\beta$ as coefficients, respectively. Some of their effects have already been analysed in the literature, but others
have not. We will consider their consequences in what follows, by analyzing  each relevant case separately.

\smallskip

\noindent
{\bf Case I: $\alpha=\beta=0$.}\, In this case
 the action (\ref{decoupling}) reduces to the Galileon action introduced by Ref.~\cite{Nicolis:2008in}. The stability conditions were studied in detail in the same reference, hence we do not repeat them here. Interestingly, it was found that the stability conditions force the speed of sound for fluctuations propagating radially to be greater than one. On the other hand, if the fourth order Galileon term ${\cal L}^{gal}_{4}$ is included, then $K_{\Omega} \ll K_{t}, K_{r}$ and the propagation of angular fluctuations is extremely sub-luminal, invalidating a quasi-static approximation.

\smallskip

\noindent
{\bf Case II: $\alpha  \neq 0$ and $\beta$=0.}
A new feature of this class of models is the disformal coupling between matter and the scalar, namely $\partial_{\mu} \pi \partial_{\nu} \pi T^{\mu \nu}$. As pointed out in Ref.~\cite{Berezhiani:2013dw}, this coupling has deep implications for the stability. Inside a matter source, the coupling introduces a kinetic term proportional to $\alpha \rho (\partial_t \varphi)^2$. If $\alpha <0$, the fluctuations behave as ghosts, forcing us to choose $\alpha >0$. In the case of massive gravity, there is only one dimensionless parameter, $\alpha\equiv 1+3\alpha_3$ (in the theory where $\beta\equiv\alpha_3+4\alpha_4=0$), leading to the so-called restricted Galileon \cite{Berezhiani:2013dw}. In this case, there is no Vainshtein solution that can be connected to the asymptotically flat solution. Around the non-flat asymptotic solution,  the radial propagation is sub-luminal, even though the extreme sub-luminarity of the angular propagation persists \cite{Berezhiani:2013dw}.

\smallskip

\noindent
{\bf Case III: $\alpha \neq 0$, $\beta \neq 0$.}
In this case, the coupling between the metric and the scalar $\hat{h}^{\mu \nu} X^{(3)}_{\mu \nu}$ gives a different picture from the previous cases and this has not been analysed so far. Inside the Vainshtein radius the solution that reduces to GR near a matter source only exists if $\beta>0$, and the solution is given by \cite{Chkareuli:2011te,Sbisa:2012zk}
\begin{equation}\label{solx0}
x_0=\pm \sqrt{\frac{\xi}{3 \beta}}.
\end{equation}
Since $x$ is constant and $A(r)\gg 1$ inside the Vainshtein radius, all functions $K_i$ in (\ref{Ks}) are dominated by  the $y\sim A(r)\gg 1$ contribution in the limit $r \ll r_V$, as shown from its definition (\ref{ydef}). In the limit $A(r) \gg 1$, we can also  ignore the coupling between the scalar and metric perturbations. By taking into account only the contributions depending on $A= (r_V/r)^3$ outside the surface of the source $r>r_s$, we find in this limit
\begin{equation}
K_t=0, \quad K_r = -12 \beta A(r) x_0, \quad K_{\Omega}= 6 \beta A(r) x_0.
\end{equation}
Thus inside the Vainshtein radius $r \ll r_V$, but outside the source surface, the speed of the fluctuations are always superluminal, for both radial and angular directions. Moreover, given the fact that $K_\Omega$ and $K_r$ have opposite signs, all solutions are unstable.

On the other hand, the instability can be avoided inside the source. For definiteness, we consider a distribution of matter given by $\rho(r)=\rho_0 (r/r_s)^s$ for $r \leq r_s$.
Inside the matter source $r \leq r_s$, the $K_i$ functions are given by
\begin{eqnarray}
K_t &=& 4 (\alpha + 3 \beta x_0)(3+s) A(r), \nonumber\\
K_r &=& - 12 \beta x_0 A(r), \quad
K_{\Omega} =  - 6 (2 +s) \beta x_0 A(r),
\end{eqnarray}
where $A(r) = \rho(r)/\left[2(3+s) \mpl \Lambda^3\right]$.
The stability and superluminality depend on the slope of the density profile, $s$, as well as on the values of the coefficients $\alpha$, $\beta$. The speeds of sound for radial and angular fluctuations read
\bea
c_r^2&=&-\frac{  12 \beta x_0 }{4 (\alpha + 3 \beta x_0)(3+s)},
\\
c_\Omega^2&=&-\frac{   6 (2 +s) \beta x_0 }{4 (\alpha + 3 \beta x_0)(3+s) },
\eea
and their ratio is $c^2_r/c^2_\Omega\,=\,2/(2+s)$. Choosing the negative branch in eq. (\ref{solx0}), $s\ge-2$,  and $\alpha$ positive and sufficiently large, both the previous expressions can be made positive avoiding instabilities.  For appropriate values of $\alpha$ and $s$, moreover, the perturbations can be made subluminal, but not so subluminal to invalidate a quasi-static approximation.

\section{Discussion:}
In this letter, we derived the effective action that describes the Vainshtein mechanism from the Horndeski action. Using the effective theory, we studied spherically symmetric solutions and perturbations around them. The solutions can be classified by two parameters $\alpha$ and $\beta$. If $\alpha \neq 0$, there appears a disformal coupling between the scalar and energy-momentum tensor. If $\beta \neq 0$, the coupling between the metric and scalar cannot be diagonalised by a local field transformation. We found that if $\beta \neq 0$, the vacuum spherically symmetric solution is pathological. Inside the Vainshtein radius, at the leading order in $(r_V/r)\gg 1$, the temporal kinetic term vanishes, leading to an extreme superluminal propagation. Moreover, the spatial kinetic term for the radial and angular modes have opposite signs, so the perturbations are unstable. This instability can be cured inside a matter source, choosing an appropriate
density profile.  In order to avoid the instability of vacuum solutions, on the other hand, we  would need to fine-tune $\beta$ such that $\beta < O((r_s/r_V)^2)$. For the case of the Earth, this condition would imply $\beta < 10^{-22}$  for a cut-off scale of order $\Lambda \sim \mpl H_0^2$. This imposes a strong restriction on the function $G_5$ appearing in the Horndeski action.

For a further understanding of this coupling mediated by $\beta\neq0$, it would be interesting to solve the system numerically when the spherical symmetry is relaxed, using the techniques developed in Ref.~\cite{Li:2013nua}. Due to the complex tensor structure of the equations, screening works very differently in non-spherically symmetric systems \cite{Hiramatsu:2012xj}. Another interesting possibility is to consider the effect of pressure, since it has already been pointed out that pressure gives contributions to $K_r$ and $K_{\Omega}$ through the disformal coupling depending on $\alpha$ \cite{Berezhiani:2013dw, Spolyar:2013maa}.

Recently, it was argued that, except for the model with only the cubic Galileon term ${\cal L}_3^{\rm gal}$, the theory looses its predictive power due to quantum corrections at ${\cal O}(1)$mm if $\Lambda$ is associated with the energy scale for the current accelerated expansion of the Universe $\Lambda \sim \mpl H_0^2$ \cite{Burrage:2012ja}. There is still a debate over whether the disformal coupling is enough to cure this problem \cite{deRham:2012ew, Berezhiani:2013dw} but we should point out that the situation might be worse when we consider a spherically symmetric vacuum solution as a background, if $\beta \neq 0$. In this case, the temporal kinetic term is of order unity $K_t \sim {\cal O}(1)$, which is much smaller than the one reduced from an order estimation $K_t \sim (r_V/r)^3$. Again the existence of matter significantly increases the kinetic term $K_t \sim (\rho/\mpl \Lambda^3)$ and the strong coupling scale depends on the matter distribution.

\acknowledgments
We thank the authors of Ref.~\cite{mixed} for sharing their preliminary results which have some overlap with our work. KK is supported by STFC grant ST/H002774/1 and ST/K0090X/1, the European Research Council and the Leverhulme trust. GN is supported by the grants PROMEP/103.5/12/3680 and CONACYT/179208. GT is supported by an STFC Advanced Fellowship ST/H005498/1.

\end{document}